\documentclass[twocolumn]{aastex61}%
\usepackage{amssymb,amsmath,amsfonts}
\usepackage{graphicx,color}
\usepackage[english]{babel}
\newcommand{\dif}{\mathrm{d}}%
\newcommand{\tdif}[2]{\frac{\dif#1}{\dif#2}}%
\makeatletter
\def\frontmatter@preabstractspace{48pt}
\makeatother

\shorttitle{Flux of high-energy cosmogenic neutrinos}
\shortauthors{Wittkowski and Kampert}

\begin{document}
\title{Predictions for the flux of high-energy cosmogenic neutrinos and the influence of the extragalactic magnetic field}

\correspondingauthor{David Wittkowski}
\email{david.wittkowski@uni-wuppertal.de}

\author{David Wittkowski}
\affiliation{Department of Physics, Bergische Universit\"at Wuppertal, Gau\ss{}stra\ss{}e 20, D-42097 Wuppertal, Germany}

\author{Karl-Heinz Kampert}
\affiliation{Department of Physics, Bergische Universit\"at Wuppertal, Gau\ss{}stra\ss{}e 20, D-42097 Wuppertal, Germany}

\begin{abstract}
Cosmogenic neutrinos originate from interactions of cosmic rays propagating through the universe with cosmic background photons. Since both high-energy cosmic rays and cosmic background photons exist, the existence of high-energy cosmogenic neutrinos is almost certain. However, their flux has not been measured so far. Therefore, we calculated the flux of high-energy cosmogenic neutrinos arriving at the Earth on the basis of elaborate four-dimensional simulations that take into account three spatial degrees of freedom and the cosmological time-evolution of the universe. Our predictions for this neutrino flux are consistent with the recent upper limits obtained from large-scale cosmic-ray experiments. We also show that the extragalactic magnetic field has a strong influence on the neutrino flux. The results of this work are important for the design of future neutrino observatories, since they allow to assess the detector volume that is necessary to detect high-energy cosmogenic neutrinos in the near future. An observation of such neutrinos would push multimessenger astronomy to hitherto unachieved energy scales.
\end{abstract}

\keywords{astroparticle physics --- cosmic rays --- magnetic fields --- neutrinos}

\section{Introduction}
Ultra-high-energy cosmic rays (UHECRs), i.e., charged nuclei with energies $E \geq 10^{18}\,\mathrm{eV}$ \citep{NaganoW2000}, are the most energetic particles detected at Earth.
Despite many years of research, the origin of UHECRs is still unknown \citep{Linsley1963}. Nevertheless, there was much progress in recent years, including the finding of indications for an extragalactic origin of UHECRs \citep{AabPAOOoal2017}.  
A promising way to get additional and independent information about the origin of UHECRs could provide the so-called ``cosmogenic neutrinos''.
These neutrinos are produced when UHECRs interact with photons from the cosmic microwave background (CMB) \citep{BeresinskyZ1969} or extragalactic background light (EBL) during the UHECRs' propagation through the universe.  
Since UHECRs and cosmic background photons from the CMB and EBL exist, we can assume the existence of high-energy cosmogenic neutrinos \citep{KaleshevkSS2002}. Up to now, they have not yet been measured though. This is likely due to too small volumes and detection times of the present observatories.

Nevertheless, large-scale projects such as the Pierre Auger Observatory and IceCube Neutrino Observatory have provided upper limits for the flux of high-energy cosmogenic neutrinos \citep{ZasPAO2017,AartsenIceCube2018}. 
On the theoretical side, one-dimensional simulations of the propagation of UHECRs, the generation of cosmogenic neutrinos, and their flight to Earth have been carried out and led to initial predictions for the neutrino flux \citep{KoteraO2011, vanVlietHB2017}. 
These simulations, however, had some significant limitations. Among them are the reduced number of spatial degrees of freedom and the fact that a structured extragalactic magnetic field, which is known to have a strong effect on the propagation of UHECRs \citep{MollerachR2013,WittkowskiDPAO2017}, cannot be taken into account in one-dimensional simulations.  

In this Letter, we present results for high-energy cosmogenic neutrinos that are based on advanced four-dimensional simulations taking into account all three spatial degrees of freedom, the cosmological time-evolution of the universe, as well as the structured extragalactic magnetic field. 
With the time-evolution of the universe, the simulations include cosmological effects such as the redshift evolution of the photon background and the adiabatic expansion of the universe.
We concentrate on neutrinos with energies $E \geq 10^{17}\,\mathrm{eV}$, since for lower energies particle interactions in the intracluster medium cannot be neglected as sources of neutrinos \citep{FangM2018}. 
Regarding the properties of the sources of UHECRs and their propagation through the universe, the simulations make use of the recently proposed astrophysical model described in \citet{WittkowskiDPAO2017,WittkowskiK2018}, which has been proven to be in very good agreement with the data from the Pierre Auger Observatory. This agreement covers the energy spectrum, chemical composition, and anisotropy in the arrival directions of the detected UHECRs.
On the basis of these simulations, we make predictions for the flux of high-energy cosmogenic neutrinos reaching the Earth, and we compare the predictions with the recent upper limits from the Pierre Auger Observatory and IceCube Neutrino Observatory. We also show that neglecting the extragalactic magnetic field strongly affects the predictions for the cosmogenic neutrino flux.

\section{\label{sec:Methods}Methods}
The four-dimensional simulations of the propagation of UHECRs and secondary neutrinos were performed using the Monte-Carlo code CRPropa 3 \citep{BatistaDEKKMSVWW2016}. We applied the astrophysical model presented in \citet{WittkowskiDPAO2017,WittkowskiK2018} and extended the simulations described in those articles towards the propagation of cosmogenic neutrinos. 

At the sources, UHECRs consisting of $^{1}\mathrm{H}$, $^{4}\mathrm{He}$, $^{14}\mathrm{N}$, $^{28}\mathrm{Si}$, and $^{56}\mathrm{Fe}$ nuclei are emitted isotropically with an energy spectrum 
\begin{equation}
\tdif{N_{0}}{E_{0}} \propto \sum_{\alpha} f_{\alpha} E_{0}^{-\gamma} \exp\!\bigg(\!-\mathcal{R}\bigg(\frac{E_{0}}{Z_{\alpha}R_{\mathrm{cut}}}-1 \bigg)\!\bigg) \,,
\end{equation} 
where $\dif N_{0}(E_{0})$ is the number of particles emitted with an energy $E\in [E_{0},E_{0}+\dif E_{0}[$. 
The parameters $f_{\alpha}$ with $\alpha\in\{\mathrm{H}, \mathrm{He}, \mathrm{N}, \mathrm{Si},\mathrm{Fe}\}$ and normalization $\sum_{\alpha} f_{\alpha} = 1$ describe the fraction of particles of an element $\alpha$ among all emitted particles and thus the chemical composition of the UHECRs at their sources.  
Furthermore, $\gamma$ is called spectral index, $\mathcal{R}(x)$ is the usual ramp function, $Z_{\alpha}$ is the atomic number of an element $\alpha$, and $R_{\mathrm{cut}}$ is the so-called cut-off rigidity.

The source parameters $f_{\alpha}$, $\gamma$, and $R_{\mathrm{cut}}$ are chosen such that the simulation results for the UHECRs arriving at the Earth are in optimal agreement with the data from the Pierre Auger Observatory. 
As has been shown in \citet{WittkowskiDPAO2017}, the optimal parameters depend on whether the influence of the EGMF on the propagation of the UHECRs is taken into account or not.  
For our main simulations, which consider the EGMF, we used the parameter values $f_{\mathrm{H}}=3.0\%$, $f_{\mathrm{He}}=2.1\%$, $f_{\mathrm{N}}=73.5\%$, $f_{\mathrm{Si}}=21.0\%$, $f_{\mathrm{Fe}}=0.4\%$, $\gamma=1.61$, and $R_{\mathrm{cut}}=10^{18.88}\,\mathrm{eV}$ \citep{WittkowskiK2018}.
In the additional simulations, where we neglected the EGMF, we used the corresponding optimal parameters 
$f_{\mathrm{H}}=11.0\%$, $f_{\mathrm{He}}=13.8\%$, $f_{\mathrm{N}}=67.9\%$, $f_{\mathrm{Si}}=7.2\%$, $f_{\mathrm{Fe}}=0.1\%$, $\gamma=0.61$, and $R_{\mathrm{cut}}=10^{18.48}\,\mathrm{eV}$ \citep{WittkowskiDPAO2017}.

The above mentioned values for the spectral index are incorporated not directly via the energy spectrum at the sources, but indirectly by a suitable post-processing of the simulation data for the UHECRs and cosmogenic neutrinos at the Earth. In order to speed up the simulations and to improve the statistics at high energies, we set the spectral index in the initial energy spectrum to one and applied the re-weighting technique described in \citet{ArmengaudSBM2007,VanVlietDis2015} to the simulation data to effectively change the spectral index to its target value. 
For an easier comparison of the obtained neutrino flux at Earth with the upper limits for the neutrino flux from the Pierre Auger Observatory and IceCube Neutrino Observatory, the simulation results for the fluxes of UHECRs and neutrinos at Earth are finally rescaled by a suitable global constant factor. This factor is chosen so that it matches the simulated flux of UHECRs at arrival energy $10^{19}\,\mathrm{eV}$ to the flux of UHECRs measured by the Pierre Auger Observatory at this energy. 

Different from the simulations in \citet{WittkowskiDPAO2017,WittkowskiK2018}, which focused on UHECRs and not neutrinos, we use here a larger maximal redshift of $z=4$, since neutrinos can reach the Earth from higher distances than UHECRs. As a further difference, we increased the radius of the spherical observer to $2\,\mathrm{Mpc}$ in order to gain sufficient statistics.

\section{\label{sec:Results}Results}
\begin{figure}[ht]
\begin{center}%
\includegraphics[width=\linewidth]{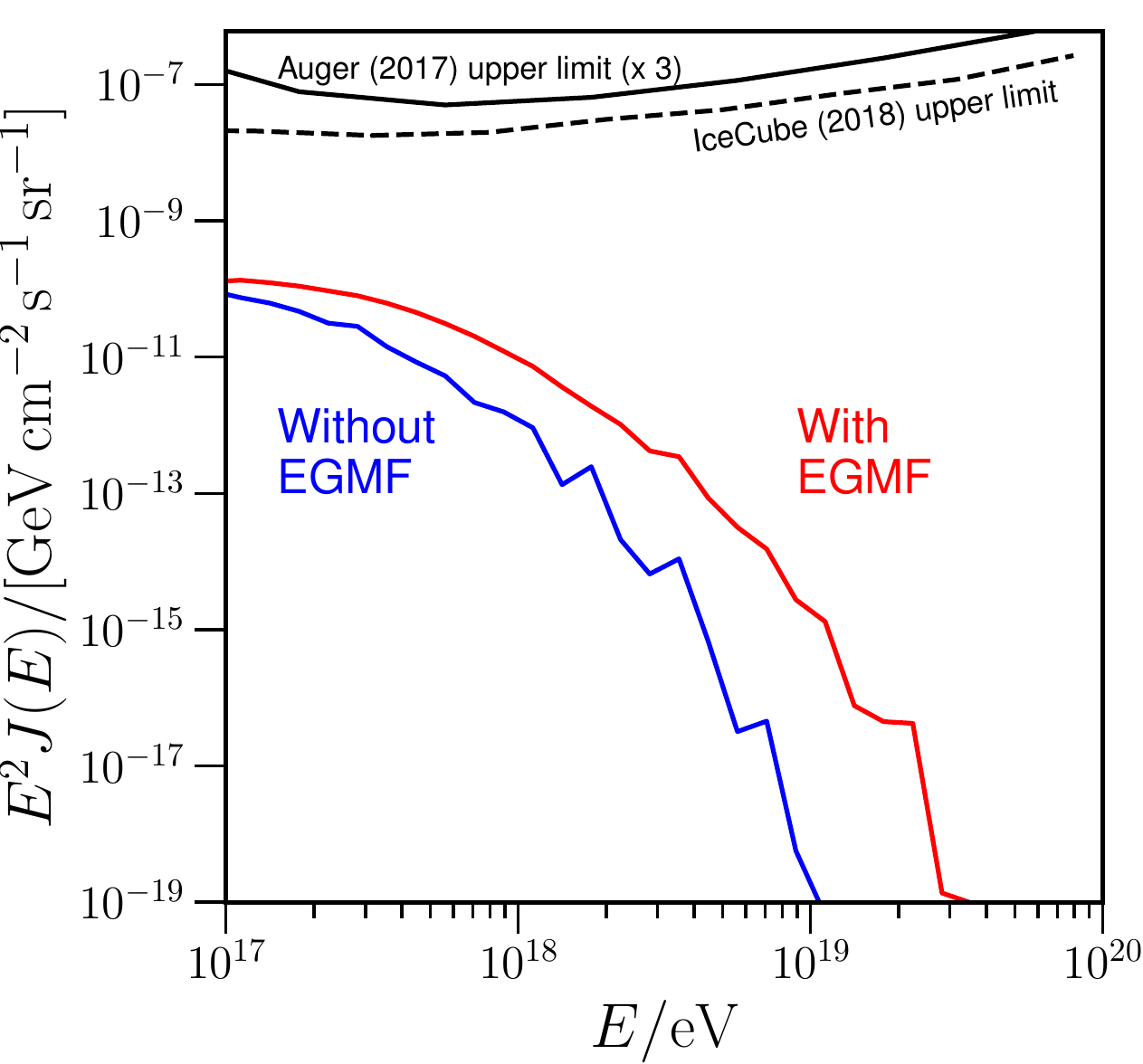}%
\caption{\label{f1}Predictions for the all-flavor flux $J(E)$ of high-energy cosmogenic neutrinos arriving at the Earth as a function of the neutrino energy $E$, when considering (red) or neglecting (blue) the EGMF. The differential upper limits (black) for the neutrino flux that were obtained from the Pierre Auger Observatory \citep{ZasPAO2017} and IceCube Neutrino Observatory \citep{AartsenIceCube2018} are shown for reference.}%
\end{center}%
\end{figure}

Figure \ref{f1} shows our predictions for the energy-dependent flux of high-energy cosmogenic neutrinos of all flavors arriving at the Earth. 
Two corresponding curves are shown for simulations taking the EGMF into account or neglecting it, respectively. 
For comparison, the figure shows also the differential upper limits for the cosmogenic neutrino flux that have been obtained from the data measured by the Pierre Auger Observatory \citep{ZasPAO2017} and the IceCube Neutrino Observatory \citep{AartsenIceCube2018}. Assuming equal ratios of neutrino flavors, the single-flavor upper limit from \citet{ZasPAO2017} has been converted to an all-flavor upper limit by multiplication by three.  

The predicted cosmogenic neutrino fluxes are several orders of magnitude below the upper limits and therefore consistent with the current measurements of the observatories. 
Since the predictions for the neutrino fluxes are much smaller than the upper limits, it is not surprising that no high-energy cosmogenic neutrinos have been detected until today. 
Both curves for the predicted cosmogenic neutrino fluxes start with approximately the same values at neutrino energy $E=10^{17}\,\mathrm{eV}$ and decrease when $E$ increases. 
The curve that is related to the presence of an EGMF is always above the other curve and the vertical distance of the curves increases with $E$. While the curves start with similar values at $E=10^{17}\,\mathrm{eV}$, the neutrino fluxes differ already by four orders of magnitude at $E=10^{19}\,\mathrm{eV}$.    
This reveals that the EGMF has a strong influence on the high-energy cosmogenic neutrino flux. 
The observation that the EGMF increases the neutrino flux can partially be explained by the fact that deflections of UHECRs in a magnetic field lead to larger path lengths and thus more interactions with the photon background.

\section{\label{sec:Conclusions}Conclusions}
Based on elaborate four-dimensional simulations, we studied the flux of cosmogenic neutrinos with energies $E \geq 10^{17}\,\mathrm{eV}$ arriving at the Earth.
For this purpose, we simulated the propagation of UHECRs through the universe, the generation of high-energy cosmogenic neutrinos by interactions of UHECRs with photons from the CMB or EBL, and the path of the neutrinos to the Earth. 
The performed simulations took into account all three spatial degrees of freedom, the cosmological time-evolution of the universe, as well as the structured EGMF.
By using an appropriate astrophysical model, a very good agreement of the simulation results and the UHECRs data from the Pierre Auger Observatory was achieved.

The simulations yielded predictions for the flux of high-energy cosmogenic neutrinos at the Earth.
These findings are consistent with the upper limits for the neutrino flux obtained by the Pierre Auger Observatory and IceCube Neutrino Observatory.
Corresponding simulations where we neglected the EGMF led to strongly different predictions with up to four orders of magnitude lower neutrino fluxes. This confirms that the EGMF has a major influence on the neutrino flux and typically must not be neglected in theoretical studies on the generation and propagation of cosmogenic neutrinos. 
Since considering an EGMF leads to higher cosmogenic neutrino fluxes, predictions from previous simulations that neglected the EGMF can be considered only as lower bounds for the cosmogenic neutrino flux.

The results of this work are important for the design of future neutrino observatories such as ARA \citep{AllisonARACollaboration2012}, ARIANNA \citep{BarwickARIANNACollaboration2015}, and GRAND \citep{MartineauHuynhBCetal2017,Fangetal2017} that are currently in the planning phase. Our predictions for the cosmogenic neutrino flux allow to assess which detector volume is necessary to detect high-energy cosmogenic neutrinos in the near future. The observation of such neutrinos is an important goal in current astrophysics, since it would push multimessenger astronomy from below $10^{15}\,\mathrm{eV}$ \citep{IceCubeMultimessenger2018} to entirely new energy scales.

\begin{acknowledgments}
\section*{Acknowledgments}
We thank the Pierre Auger Collaboration for helpful discussions. The simulations were carried out on the pleiades cluster at the University of Wuppertal, which was supported by the Deutsche Forschungsgemeinschaft (DFG). Further financial support by the BMBF Verbundforschung Astroteilchenphysik is acknowledged.
\end{acknowledgments}

\end{document}